\documentclass[useAMS]{mn2e}
\usepackage{graphicx}
\usepackage{color}

\begin{document}
\title[MOST photometry of the RRd Lyrae variable AQ Leo]{MOST\thanks{MOST is a Canadian 
Space Agency mission, jointly operated by Dynacon Inc., the University of Toronto Institute for 
Aerospace Studies and the  University of British Columbia, with the assistance of the University of 
Vienna.} photometry of the RRd Lyrae variable AQ Leo: Two radial modes, 32 combination frequencies,
and beyond}

\author[M. Gruberbauer, K. Kolenberg, et al.]{Michael Gruberbauer$^{1}$,  Katrien Kolenberg$^{1,3}$,
Jason F. Rowe$^{2}$, Daniel Huber$^{1}$,
\and Jaymie M. Matthews$^{2}$, Piet Reegen$^{1}$, Rainer Kuschnig$^{2}$, Chris Cameron$^{2}$, 
\and Thomas Kallinger$^{1}$, Werner W. Weiss$^{1}$, David B. Guenther$^{4}$, Anthony F. J. 
Moffat$^{5}$, 
\and Slavek Rucinski$^{6}$, Dimitar Sasselov$^{7}$, Gordon A. H. Walker$^{2}$ \\
\\
$^1$Institute of Astronomy, University of Vienna, T\"urkenschanzstrasse 17, A-1180 Vienna, Austria
\\
$^2$Department of Physics \& Astronomy, University of British Columbia, 6224 Agtricultural Road, 
Vancouver, B. C., V6T 1Z1, Canada 
\\
$^3$Institute of Astronomy, University of Louvain, Celestijnenlaan 200D, B-3001 Heverlee, Belgium
\\
$^4$Department of Astronomy and Physics, St. Mary's University, Halifax, Nova Scotia, NS B3H 3C3, Canada
\\
$^5$D\'epartement de physique, Universit\'e de Montr\'eal, Montr\'eal, Queb\'ec, QC H3C 3J7, Canada
\\
$^6$David Dunlap Observatory, Department of Astronomy, University of Toronto, Toronto, Ontario, ON L4C 4Y6,
Canada
\\
$^7$Harvard-Smithsonian Center for Astrophysics, Cambridge, Massachusetts, MA 02138, USA
}

\date{Accepted 2007 May 30. Received 2007 May 30; in original form  2007 April 16; minor revision 2007 July 9}

\maketitle

\begin{abstract}
Highly precise and nearly uninterrupted optical photometry of the RR Lyrae star AQ Leo was obtained 
with the MOST (Microvariability \& Oscillations of STars) satellite over 34.4 days in February-March 
2005.  AQ Leo was the first known double-mode RR Lyrae pulsator (RRd star).  Three decades after 
its discovery, MOST observations have revealed that AQ Leo oscillates with at least 42 frequencies, 
of which 32 are linear combinations (up to the sixth order) of the radial fundamental mode and its first 
overtone. Evidence for period changes of these modes is found in the data. The other intrinsic frequencies 
may represent an additional nonradial pulsation mode and its harmonics (plus linear combinations) 
which warrant theoretical modeling. The unprecedented number of frequencies detected with amplitudes 
down to millimag precision also presents an opportunity to test nonlinear theories of mode growth and 
saturation in RR Lyrae pulsators.
\end{abstract}


\section{Introduction}

The human part of the history of AQ Leo ($V \sim 12.6$) begins with its classification as a variable of 
the RW Aurigae type by Hoffmeister (1944) at the Sonneberg Observatory. Fiftteen years later, Wenzel 
(1961) concluded that the star is an RR Lyrae pulsator with a strongly variable light curve. Jerzykiewicz 
\& Wenzel (1977) thoroughly analysed their photometry gathered during 1973 - 1975 to recognise for the 
first time that AQ Leo is a double-mode RR Lyrae star. They identified the prototype of the class. 
Jerzykiewicz, Schult \& Wenzel (1982) later re-evaluated the data and their comparisons to earlier 
archived exposures raised the possibility of a period change in the early 1970s. In the meantime, 
double-mode RR Lyraes, designated RRd stars, have been identified in globular clusters, dwarf spheroidal 
galaxies, the Magellanic Clouds and the Galactic field (see, e.g., Clement et al. 1991).

RRd stars pulsate in a combination of first overtone and fundamental radial modes. The ratio of the first
overtone to the fundamental period is about 0.74 - 0.75 (see, e.g., Alcock et al. 2000). Usually the 
amplitude of the first overtone is stronger, but not always so (e.g., Oaster et al. 2006). RRd variables 
are particularly important because their double-mode nature affords an opportunity to determine their 
masses based on the period ratio, largely independently of stellar evolution theory, and hence to a much 
higher precision than is possible for monomode RR Lyrae stars (Kov\`{a}cs et al 1991).

AQ Leo happens to fall in the field around a MOST space mission Primary Science Target, $\iota$ Leo. This 
alignment presented an ideal opportunity to obtain the first new light curve of this star in three decades, 
with very complete time coverage and unequalled photometric precision. The long time baseline between  
the JW77 photometry and the MOST observations is well suited to search for period changes in AQ Leo.  
The precision of the MOST photometry means that the amplitude ratio of the two modes can be measured 
more accurately than before; however, in the MOST custom passband, not in a standard filter system. If 
AQ Leo is reobserved by MOST in the future, it will be possible to test theoretical predictions of mode 
switching in RRd stars.  The sampling and precision of the MOST photometry also make it possible to 
search for {\em nonradial} modes in an RRd star with a sensitivity that cannot be achieved in groundbased 
data.

\section{MOST photometry}

The MOST (Microvariability \& Oscillations of STars) space mission was designed to obtain high-precision
optical photometry of bright stars, monitored for weeks at a time with high duty cycle and time sampling of
at least once per minute (Walker et al. 2003).  MOST is a microsatellite 
housing a 15-cm Maksutov telescope feeding a CCD photometer through a single custom broadband filter
(350 - 700 nm).  Its polar Sun-synchronous orbit allows it to observe stars in its Continuous Viewing Zone 
for up to two months without interruption.

AQ Leo was observed by MOST satellite in the same field as its Primary Science Target $\iota$ Leo, for 34.434 
days during 14 February -- 22 March 2005 (HJD 2453416.671 -- 2453451.105).  One of the faintest stars
ever observed by MOST at $V \sim 12.6$, AQ Leo was a Direct Imaging target. Direct Imaging is the MOST 
observing mode where a defocused star image is projected onto an open area of the Science CCD (see Rowe 
et al. 2006). The star is sampled within a $20 \times 20$-pixel subraster of the CCD and photometry is 
extracted through aperture methods and PSF (Point Spread Function) fitting of the image profile. The 
exposure time was 25 s and exposures were made every 30 s. The data were reduced independently by two 
of the authors, JFR at UBC and DH at IfA Vienna, and the results then compared.  

The UBC MOST Direct Imaging photometry reduction pipeline is similar to the treatment of groundbased CCD
photometry in that it measures the stellar flux though a combination of an aperture of pixels centred on the 
star and fitting of its Point Spread Function (PSF) image profile.  To minimise the effects of pointing errors on 
the photometry, dark and flatfield corrections were performed by monitoring individual pixel responses during 
test exposures on fields empty of stars brighter than the background.  The correlation in the raw photometry 
between the instrumental magnitude light curve and the estimated sky background was removed as described 
in Rowe et al. (2006).  

The IfA Vienna reduction was based on a pixel-to-pixel decorrelation technique developed by Reegen et al. 
(2006) for MOST Fabry Imaging photometry.  To define a fixed aperture mask by which pixels illuminated by the 
star and by the background can be identified, the PSFs of the star on all subraster images must be aligned.  
This is not the case in the raw images due to minor pointing errors during the observing run. As a first step, the 
centroid position of the stellar PSF for each image is computed. Subsequently, all other pixel intensities relative 
to the centroid coordinates are computed by 2-dimensional linear interpolation. Finally, all pixels of each image 
are shifted by integer values in x and y to a pre-defined centre position. Since this shifting process decreases 
the resulting image size common to all exposures, frames suffering from pointing errors larger than 3 pixels  
(about 9 arcsec) from the subraster centre are rejected. (This pointing threshold eliminated 7\% of all the 
exposures in the raw AQ Leo time series.)  Once all the images are aligned, the decorrelation can be computed 
to reduce the background effects of modulated stray light.

\begin{figure}
\resizebox{\hsize}{!}
{\includegraphics{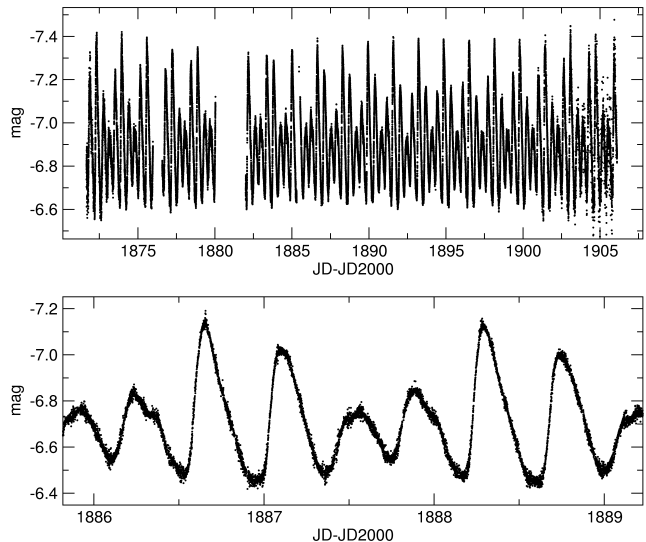}}
 \caption{The light curve of AQ Leo obtained by MOST. The entire time series (upper panel), as well as an 
expanded view of 3.7 days of the light curve (lower panel).}
\label{fig:lightcurve}
\end{figure}

Both reductions led to consistent light curves and frequency identifications.  The IfA light curve has higher 
point-to-point scatter than the UBC reduction, and the UBC reduction is presented here as the AQ Leo light 
curve.  However, both light curves were searched independently by the frequency analysis techniques described
in the next section, and the results are consistent for both reductions.  The final duty cycle of the reduced 
photometry is 82\%.  The reduced light curve is presented in Figure \ref{fig:lightcurve}.  The AQ Leo raw data and 
reduced light curve are available in the MOST Public Data Archive, accessed via the Science page of the MOST
web site: {\sf www.astro.ubc.ca/MOST}.

\section{Frequency analysis}

Frequency analyses of the MOST AQ Leo light curve were performed independently with three different routines: 
Period04 (Lenz \& Breger 2005), SigSpec (Reegen 2007) and nonlinear least squares fitting and bootstrapping 
(Rowe et al. 2006b; see also Cameron et al. 2006) and the results compared to ensure that only significant 
intrinsic stellar frequencies were identified. 

\subsection{Period04}

Period04 (Lenz \& Breger 2005) is a routine which applies a single-frequency power spectrum and simultaneous 
multi-frequency sine-wave fitting. The program includes advanced options such as the calculation of optimal light 
curve fits for multi-periodic signals including harmonics, combination frequencies, and combs of equally spaced 
frequencies. In this analysis, we identified frequencies through successive prewhitening. As each new significant 
frequency was added to the fit, the optimum values of first overtone frequency $f_1$ and the fundamental $f_0$, and 
their amplitudes and phases, were determined by minimising the residual rms error of the fit. After no additional 
frequencies were detected in the data, we applied a simultaneous least-squares fit of all the independent 
frequencies and their significant linear combinations, as well as all amplitudes and phases.  This included a final 
optimisation of the first overtone and radial fundamental frequencies. 

Besides determining the uncertainties from the error matrix of a least-squares calculation and/or from analytically 
derived formulae, Period04 also provides an interface for estimating the uncertainties of the fit parameters by means 
of Monte Carlo simulations. A set of time series is generated for which the times are the same as for the original 
time series, and for which the magnitudes (intensities) are calculated from those predicted by the last fit plus 
Gaussian noise.  A least-squares calculation is made for every time string, and the resulting uncertainties are 
based on the distribution of the fit parameters.

\subsection{Nonlinear least-squares fitting and bootstrapping}

The nonlinear least-squares approach we used for determining the sinusoid parameters from the AQ Leo time 
series follows the philosophy of Period04 (Lenz \& Breger 2005) described above. Discrete Fourier Transforms 
(DFTs) are calculated and an updated fit is subtracted from the data successively, until there is no meaningful 
change in the fit residuals.  The data are fit using an equation of the form
\begin{equation}
mag = A_0 + \sum _{j=1,n} A_j \cos (2\pi f_j t - \theta_j),
\end{equation}
where $A_0$ is a linear offset and $f_j$, $A_j$ and $\theta_j$ are the frequency, amplitude and phase for each 
successive peak found in the amplitude spectrum. The uncertainties calculated for time series parameters 
derived from nonlinear least-squares fitting depend on the noise of the data (which may be a combination of 
instrumental and random processes) and on the time sampling. 

It is also known that fitted phase and frequency parameters are correlated, leading to underestimated 
uncertainties in these parameters when calculated from a covariance matrix (see, e.g., Montgomery \& 
O'Donoghue 1999).  The "bootstrap" (see Wall \& Jenkins 2003) is a very effective way to assess the 
uncertainties in these fitted parameters. The procedure has recently been used in a number of MOST 
applications (see, e.g., Cameron et al. 2006, Rowe et al. 2006b, Saio et al. 2006).  Clement et al. (1992) 
also used a bootstrap to estimate the uncertainties in Fourier parameters they derived for RR Lyrae stars. 
Bootstrapping produces a distribution for each calculated parameter by constructing a large number of 
light curves from the original data. No assumptions must be made about noise properties of the data and
individual photometric errors are not required for the calculations. Each new light curve is assembled by
randomly selecting $N$ points from the original light curve (also containing $N$ points) with the possibility 
of replacement. The new synthetic light curves preserve the noise properties of the original data. The fit is 
repeated for each new light curve, eventually building distributions in each of the fit parameters. We then
estimate the 1$\sigma$ error bars from the analytic expression for the standard deviation of each distribution
under the assumption that they are normally distributed. Each distribution is checked to ensure this
assumption is valid.  The uncertainties are listed in Table 1 and give consistent answers with SigSpec, as 
described below. We point out that bootstrapping only estimates the uncertainties in parameters.  It does 
not refine the parameter values or assign significances to them.

\subsection{SigSpec and background correction}

A  third independent frequency search was performed, where the statistical significances of peaks in the DFT 
of the AQ Leo light curve were estimated through SigSpec (Reegen 2007).  This routine calculates an unbiased 
False Alarm Probability (based on white noise) associated with the amplitude of a peak in Fourier space. It 
produces a significance spectrum containing all frequencies above a certain threshold in significance, as 
determined by successive prewhitening. 

The analyses above identify frequencies present in the data with high statistical significance and signal-to-noise,
but some of these may be due to background variations in the data, not intrinsic to the star. The frequencies of 
some background artifacts in MOST data are known; e.g., the orbital frequency of the satellite and its harmonics,
at which stray light modulation is observed.  We carried out tests to distinguish periodic effects in the background 
from intrinsic stellar signal.  We did this by analysing not only the variation of the mean integrated intensity within 
an aperture surrounding the star ("stellar signal", which includes the background in those pixels), but also the 
mean integrated intensity outside a larger aperture which includes the outermost pixels of the CCD subraster 
("background signal"). Since our three independent frequency analyses yield consistent results, we chose to use
the list of frequencies, amplitudes and phases from SigSpec to compare the significance spectra of the stellar 
signal light curve and the background signal light curve. 

First, peaks from the stellar signal DFT and the background DFT are flagged if their frequencies agree within 
the uncertainty. Based on extensive numerical simulations, Kallinge, Reegen \& Weiss (2007) found the upper limit for the 
frequency uncertainty in a time series to be
\begin{center}
\begin{equation} \label{kalres}
\Delta f = \frac{1}{\Delta T * \sqrt{sig(A,f,\theta)}}
\end{equation}
\end{center}
where $\Delta T$ is the total data set length in days and $sig(A,f,\theta)$ is the spectral significance for a given 
amplitude, frequency, and phase, as defined in Reegen (2007). This upper limit on frequency uncertainty is more 
conservative ($\approx$ 4 times larger) than the value for the frequency error estimated by Montgomery \& 
O'Donoghue (1999), based on an analytical solution for the error of a least--squares fit of a sinusoidal signal. If 
more than one coincident background peak is found around a given stellar peak within the frequency resolution, 
then the background peak with the highest significance is used for comparison to obtain the most conservative 
(and presumably the safest) solution.

Next, assuming stray-light induced artifacts are additive in terms of their intensities, the peak amplitudes found to 
be significant in the background are rescaled from the mean background signal level to the mean stellar signal 
level in order to make the relative amplitudes (and hence the significances) comparable. Now, with the amplitude 
and spectral significance of a stellar Fourier peak and the rescaled amplitude and spectral significance of its 
coincident background peak, the "conditional probability" that the stellar peak is not induced by background is 
calculated.  The False Alarm Probability is transformed into a conditional spectral significance, and only those 
stellar peaks for which this is above $5.46$ (corresponding roughly to an amplitude ratio of 4 for the target peak 
and the rescaled background peak) survive the cut. A detailed publication on this method, dubbed CINDERELLA 
(Comparison of INDEpendent RELative Least-squares Amplitudes), is in preparation (Reegen et al. 2007).

Uncertainties in frequency, amplitude and phase were first estimated through the relations of Montgomery \& 
O'Donoghue (1999) and also approximated according to Kallinger et al. (2007). The resulting uncertainties are
consistent with the bootstrap values described above, except where two frequencies are close to the resolution 
limit.  In these cases, which apply to only 4 pairs of identified frequencies, the least-squares fitting can be 
perturbed, adding further small systematic errors added to the fitted values.  

\subsection{The largest intrinsic variations observed by MOST - an unexpected challenge}

MOST was designed to detect and characterise very low-amplitude stellar oscillations, down to a few 
$\mu$mag.  AQ Leo is an exception to the MOST target list, varying over a range of a few $\times$ 0.1 mag.
A common MOST data reduction procedure of correcting the data is to apply a running mean to reduce the 
MOST-orbit-modulated stray light signal.  In this case, however, the stellar variability is large enough that this 
introduces a modulation because the relative contribution of the stray light variation changes with the mean 
brightness of the star. By fitting sky values versus instrumental magnitude, the effects of straylight can 
be largely removed, but the intrinsic stellar variation serves to modulate the relative contribution of the stray 
light to the light curve. This introduces a small artifact signal sidelobe mirroring the MOST orbital frequency 
but reduced in amplitude compared to the intrinsic stellar signal by a factor of about 100. For example, the 
frequency $f_1$ in Table \ref{tab:freqs} will be reproduced at a frequency equal to the MOST orbital frequency 
minus $f_1$, but at a much lower amplitude.

We tested this through simulations by mimicking the raw stellar photometry by adding a constant offset to
the background light curve and subsequently introducing the undeniably intrinsic stellar variability represented 
by the linear combinations of the first overtone $f_1$ and fundamental mode $f_0$ of AQ Leo. As suspected, in 
addition to the background frequencies and the aforementioned linear combinations, peaks due to beat 
frequencies among $f_1$, $f_0$ and the MOST orbital period appear in the amplitude spectrum.  These 
frequencies are recognised and rejected in the analyses of versions of the light curve reduced with a running 
mean, and are not seen in reductions which do not apply this filtering technique.  The MOST team is now alert
to this effect for any possible future observations of large-amplitude variable stars.

\subsection{Intrinsic frequencies in AQ Leo}

As a result of the described procedures, we identified 42 frequencies in the AQ Leo time series intrinsic to the 
star.  All the approaches described above led to this consistent solution. The two dominant frequencies are the 
first radial overtone $f_1$ and the fundamental radial mode $f_0$.  Of the remaining frequencies, 32 are linear 
combinations of these two up to the 6th order (see the discussion in the next section). The uncertainties of the 
frequencies, amplitudes and phases were estimated through extensive Monte Carlo simulations as well as 
through bootstrapping, as described above.  The resulting uncertainties are comparable with those obtained 
using the formulation of Montgomery \& O'Donoghue (1999). All 42 frequencies pass the conditional spectral 
significance test and cannot be reproduced from the background time series which contains solely instrumental 
signal and stray light.   

The 42 frequencies include 11 which are not linear combinations of  $f_0$ and $f_1$. These are 1.96 d$^{-1}$, 
one which corresponds to its first harmonic at 3.92 d$^{-1}$ (and which has a higher amplitude), and their 
linear combinations with $f_0$ and $f_1$.  We were very cautious about these frequencies because of their 
proximity to integer numbers of cycles per day.  The nearly continuous MOST time series does not suffer from 
cycle/day aliases found in groundbased single-site data due to to night/day gaps, but there are subtle 
modulations of stray light at 1, 2, 3 and 4 d$^{-1}$ due to MOST's Sun-synchronous orbit (which brings the 
satellite to nearly the same point over the Earth after 1 day). However, our analysis identified and rejected 
those artifacts at 2.00 and 4.00 d$^{-1}$ present in the background measurements and did not reject the 
frequencies at 1.96 and 3.92 d$^{-1}$, which differ from potential artifacts by more than the frequency
resolution of the data set.  Indeed, these frequencies are retained with high significance and signal-to-noise.
To stress their importance among the set of 42 frequencies, we hereafter refer to them as $f_i = 1.96\,d^{-1}$ 
and $f_{ii} = 3.92\,d^{-1}$. 

\begin{figure}
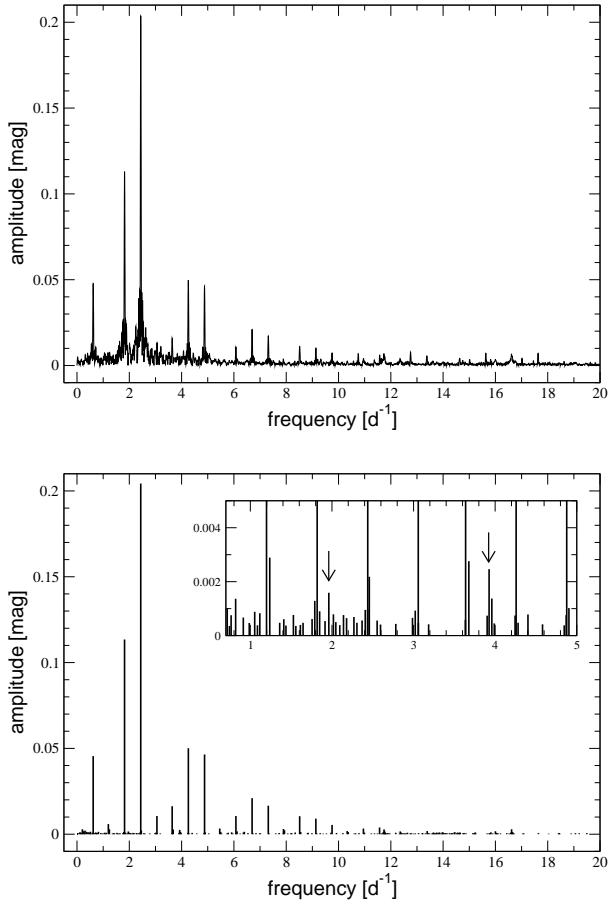

\includegraphics[width=8cm]{figure2a.eps}
\\[0.5cm]
\includegraphics[trim = 0cm 0cm 0cm  0.0cm, width=8cm,clip=true]{figure2b.eps}
\caption{{\it Upper panel} : The amplitude spectrum of the AQ Leo light curve.  Most of the linear combinations 
of $f_0$ and $f_1$ can be identified by eye. {\it Lower Panel} : The amplitude spectrum for frequencies identified 
by SigSpec. The inset is a close-up of the low frequency range containing $f_i$ and $f_{ii}$.}
\label{fig:rawspec}
\end{figure}

The frequencies, their amplitudes, phases and uncertainties are listed in Table \ref{tab:freqs}.  The amplitude
spectrum showing these frequencies is plotted in Figure \ref{fig:rawspec}.

\begin{table*} 
\caption{Frequencies identified in AQ Leo from the MOST photometry by SigSpec, which are consistent 
with the other frequency analysis techniques described in the text.  Amplitudes, phases, significances, 
and uncertainties (derived by bootstrapping) are also listed. The phases correspond to epoch JD2000 = 
JD2451545.}
  \label{tab:freqs}
  \em id: \rm identification,
  \em f: \rm frequency [d$^{-1}$],
  \em ${\sigma}_f$: \rm error in frequency [d$^{-1}$] (1$\sigma$),
  \em A: \rm amplitude [mmag],
  \em ${\sigma}_A$: \rm error in amplitude [mmag] (1$\sigma$),
  \em sig.: \rm spectral significance, 
  \em $\theta$: \rm phase,
  \em ${\sigma}_{\theta}$: \rm error in phase (1$\sigma$),
  \em comb.err.: \rm deviation $f-f_c$ from the numerical value for combinations $f_c=af_0+bf_1+cf_i$ 
  \begin{center}
  \begin{tabular}{c c c r r r r r r r r}
  \hline
  & \bf{id} & & \bf{f} & \bf{${\sigma}_{\rm f}$} & \bf{A} & \bf{${\sigma}_{\rm_A}$} & \bf{sig.} & \bf{$\theta$} & \bf{${\sigma}_{\theta}$} & 
\bf{comb.err.}\\
  \hline
  
{$-$} & {$f_1$} & {$-$} &  {$2.43821$} 		& {$0.00001$}& {$204.3$}		&{$0.106$} & {$11982$} & {$-1.855$} 	&{$0.001$} & {$0.00000$}\\
{$-$} & {$2f_1$} & {$-$} & {$4.87673$}		& {$0.00004$}& {$46.4$}			&{$0.102$} & {$5931$} & {$1.246$}		&{$0.005$} & {$0.00031$}\\
{$-$} & {$3f_1$} & {$-$} &{$7.31515$}		& {$0.00013$}& {$16.6$}			&{$0.101$} & {$2502$}& {$-2.417$}		&{$0.014$} & {$0.00052$}\\
{$-$} &{$4f_1$}&{$-$}& {$9.75228$}    		& {$0.00064$}& {$5.4$}		&{$0.118$} & {$521$}& {$-2.855$}		&{$0.069$} & {$-0.00056$}\\
{$-$} &{$5f_1$}&{$-$}& {$12.18890$}  		& {$0.00188$}& {$0.8$}		&{$0.103$} & {$16$}&  {$2.344$}		&{$0.141$} & {$-0.00215$}\\
{$f_0$} & {$-$} & {$-$} &{$1.81896$}    		& {$0.00002$}& {$113.5$}		&{$0.098$} & {$10702$}&  {$1.612$}		&{$0.002$} & {$0.00000$}\\
{$2f_0$} & {$-$} & {$-$} & {$3.63822$} 		& {$0.00019$}& {$16.3$}		&{$0.105$} & {$2780$}&  {$2.777$}		&{$0.022$} & {$0.00030$}\\
{$3f_0$} &{$-$}&{$-$}& {$5.45637$}     		& {$0.00098$}& {$3.3$}		&{$0.120$} & {$203$}& {$2.703$}		&{$0.113$} & {$-0.00051$}\\
{$f_0$} & {$f_1$} & {$-$} & {$4.25723$}		& {$0.00004$}& {$50.1$}		&{$0.099$} & {$5001$} & {$2.596$}		&{$0.005$} & {$0.00006$}\\
{$-f_0$} & {$f_1$} & {$-$} &{$0.61923$}		& {$0.00005$}& {$45.5$}		&{$0.120$} & {$8321$} & {$-1.097$}		&{$0.006$} & {$-0.00002$}\\
{$f_0$} & {$2f_1$} & {$-$} &{$6.69543$}		& {$0.00010$}& {$21.0$}		&{$0.104$} & {$3268$} & {$2.507$}		&{$0.011$} & {$0.00005$}\\
{$-f_0$} &{$2f_1$}&{$-$}& {$3.05758$}  		& {$0.00034$}& {$10.6$}		&{$0.108$} & {$1392$} & {$1.245$}		&{$0.038$} & {$0.00012$}\\
{$f_0$} &{$3f_1$}&{$-$}& {$9.13385$}			& {$0.00025$}& {$9.0$}		&{$0.103$} & {$1304$} & {$-2.346$}		&{$0.028$} & {$0.00026$}\\
{$-f_0$} &{$3f_1$}&{$-$}& {$5.49405$} 		& {$0.00288$}& {$1.3$}		&{$0.115$} & {$38$} & {$1.066$}		&{$0.305$} & {$-0.00162$}\\
{$f_0$} &{$4f_1$}&{$-$}& {$11.56975$} 		& {$0.00053$}& {$3.9$}		&{$0.105$} & {$275$} & {$1.323$}		&{$0.059$} & {$-0.00205$}\\
{$-f_0$} &{$4f_1$}&{$-$}& {$7.93297$}		& {$0.00103$}& {$2.3$}		&{$0.115$} & {$104$} & {$-2.354$}		&{$0.129$} & {$-0.00091$}\\
{$-f_0$} &{$5f_1$}&{$-$}& {$10.37055$}		& {$0.00225$}& {$1.2$}		&{$0.104$} & {$32$} & {$3.033$}		&{$0.210$} & {$-0.00154$}\\
{$2f_0$} &{$f_1$}&{$-$}& {$6.07720$}   		& {$0.00021$}& {$10.6$}		&{$0.102$} & {$1503$} & {$-0.727$}		&{$0.025$} & {$0.00107$}\\
{$2f_0$} &{$-f_1$}&{$-$}& {$1.19802$}  		& {$0.00039$}& {$5.9$}		&{$0.107$} & {$604$} & {$2.930$}		&{$0.048$} & {$-0.00169$}\\
{$2f_0$} &{$2f_1$}&{$-$}& {$8.51455$} 		& {$0.00023$}& {$10.5$}		&{$0.103$} & {$1606$} & {$1.244$}		&{$0.024$} & {$0.00021$}\\
{$-2f_0$} &{$2f_1$}&{$-$}& {$1.23813$}		& {$0.00087$}& {$2.9$}		&{$0.105$} & {$161$} & {$2.125$}		&{$0.102$} & {$-0.00037$}\\
{$2f_0$} &{$3f_1$}&{$-$}& {$10.95270$}		& {$0.00099$}& {$3.3$}		&{$0.104$} & {$203$} & {$-0.148$}		&{$0.114$} & {$0.00015$}\\
{$-2f_0$} &{$3f_1$}&{$-$}& {$3.67725$}		& {$0.00114$}& {$2.8$}		&{$0.108$} & {$152$} & {$1.295$}		&{$0.133$} & {$0.00054$}\\
{$2f_0$} &{$4f_1$}&{$-$}& {$13.38950$}		& {$0.00097$}& {$1.8$}		&{$0.105$} & {$67$} & {$1.820$}		&{$0.120$} & {$-0.00126$}\\
{$-2f_0$} &{$4f_1$}&{$-$}& {$6.12023$}		& {$0.00137$}& {$1.7$}		&{$0.106$} & {$57$} & {$1.765$}		&{$0.148$} & {$0.00531$}\\
{$2f_0$} &{$5f_1$}&{$-$}& {$15.82902$}		& {$0.00170$}& {$1.1$}		&{$0.103$} & {$26$} & {$-2.082$}		&{$0.179$} & {$0.00005$}\\
{$-2f_0$} &{$5f_1$}&{$-$}& {$8.55350$}		& {$0.00189$}& {$1.3$}		&{$0.101$} & {$34$} & {$0.463$}		&{$0.209$} & {$0.00037$}\\
{$-2f_0$} &{$6f_1$}&{$-$}& {$10.99215$}		& {$0.00366$}& {$0.9$}		&{$0.101$} & {$19$} & {$-0.578$}		&{$0.322$} & {$0.00081$}\\
{$3f_0$} &{$f_1$}&{$-$}& {$7.89427$}    		& {$0.00093$}& {$3.0$}		&{$0.113$} & {$173$} & {$-0.447$}		&{$0.117$} & {$-0.00082$}\\
{$3f_0$} &{$2f_1$}&{$-$}& {$10.33075$}		& {$0.00155$}& {$1.8$}		&{$0.104$} & {$68$} & {$2.806$}		&{$0.168$} & {$-0.00255$}\\
{$3f_0$} &{$3f_1$}&{$-$}& {$12.77310$}		& {$0.00410$}& {$0.7$}		&{$0.102$} & {$11$} & {$2.485$}		&{$0.272$} & {$0.00159$}\\
{$4f_0$} &{$f_1$}&{$-$}& {$9.71217$}     		& {$0.00329$}& {$0.9$}		&{$0.117$} & {$17$} & {$1.952$}		&{$0.351$} & {$-0.00188$}\\
{$-$} &{$-$}&{$f_i$}& {$1.96161$}       		& {$0.00257$}& {$1.6$}		&{$0.100$} & {$53$}&  {$1.114$}		&{$0.381$} & {$0.00000$}\\
{$-$} &{$f_1$}&{$f_i$}& {$4.40172$}      		& {$0.00229$}& {$0.8$}		&{$0.100$} & {$14$}&  {$-0.942$}		&{$0.292$} & {$0.00190$}\\
{$-$} &{$-$}&{$2f_i=f_{ii}$}& {$3.92593$}		& {$0.00144$}& {$2.5$}		&{$0.117$} & {$122$}& {$2.014$}		&{$0.117$} & {$0.00271$}\\
{$-$} &{$f_1$}&{$2f_i$}& {$6.36320$}		& {$0.00240$}& {$1.0$}		&{$0.109$} & {$23$}& {$-2.042$}		&{$0.216$} & {$0.00177$}\\
{$f0$} &{$f_1$}&{$2f_i$}& {$8.18505$}		& {$0.00246$}& {$0.7$}		&{$0.103$} & {$11$}& {$-2.445$}		&{$0.289$} & {$0.00466$}\\
{$-3f0$} &{$3f_1$}&{$2f_i$}& {$5.78067$}		& {$0.00238$}& {$0.8$}		&{$0.101$} & {$16$} & {$-1.855$}		&{$0.319$} & {$-0.00030$}\\
{$-4f0$} &{$2f_1$}&{$2f_i$}& {$1.52777$}		& {$0.00382$}& {$0.8$}		&{$0.101$} & {$13$} & {$-1.957$}		&{$0.248$} & {$0.00397$}\\
{$-4f0$} & {$3f_1$}&{$2f_i$}&{$3.95930$}		& {$0.00167$}& {$1.4$}		&{$0.124$} & {$40$} & {$-2.055$}		&{$0.136$} & {$-0.00271$}\\
{$-4f0$} & {$4f_1$}&{$2f_i$}&{$6.39875$}		& {$0.00193$}& {$1.4$}		&{$0.109$} & {$43$} & {$0.931$}		&{$0.169$} & {$-0.00147$}\\
{$-4f0$} &{$5f_1$}&{$2f_i$}& {$8.83928$}		& {$0.00445$}& {$0.8$}		&{$0.101$} & {$15$} & {$-2.741$}		&{$0.632$} & {$0.00085$}\\

\hline
\end{tabular}  
\end{center}
\end{table*}

\section{Discussion}

\subsection{Combination frequencies of the two dominant modes}

The MOST photometry has revealed an unprecedented number of frequencies (42) in AQ Leo. Jerzykiewicz 
\& Wenzel (1977) (hereafter JW77) first detected the first radial overtone and fundamental radial modes in 
this star. They modeled their light curve with a double-harmonic series and by fitting different numbers of 
higher-order terms until the standard deviation of the fitted curve was near the lower limit of the mean error 
of a single observation.  JW77 arrived at a fit with 14 frequencies, of which 12 are combination frequencies 
of the two dominant modes.  Of our 42 frequencies, 32 are linear combinations of those same modes.  

\begin{figure}
\includegraphics[width=8cm]{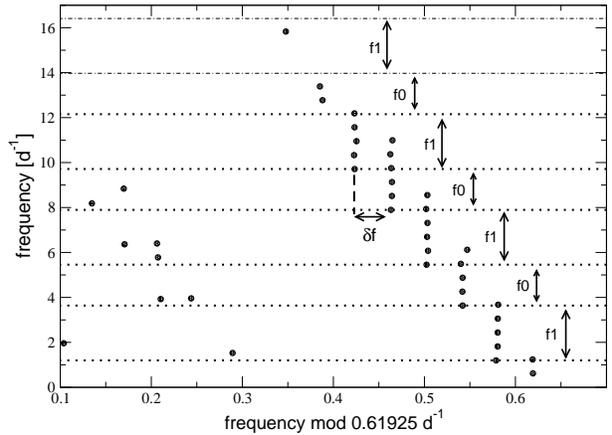}
 \caption{Graphical presentations of the linear combinations of the two dominant modes in AQ Leo, in the 
form of "echelle" diagrams.  The observed frequencies folded at the beat frequency $f_1 - f_0$ = 0.61925  
d$^{-1}$ show differences of ${\delta}f = 3{\cdot} f_1 - 4{\cdot}f_0$ in a regular pattern alternating by $f_0$ 
and $f_1$.}  
\label{fig:jumps}
\end{figure}

The numerical combinations can be obtained arithmetically in Table \ref{tab:freqs} but are much easier to see 
when the frequencies are plotted in "\'echelle" diagrams like Figure \ref{fig:jumps}.  These diagrams have the 
same form as the \'echelle diagrams used in asteroseismology to highlight the asymptotic distribution of 
nonradial $p$-modes.  The frequencies are folded modulo a common frequency spacing on the x-axis and 
plotted at their original values on the y-axis.  In the case of the MOST AQ Leo frequencies (and for RRd 
frequencies in general), the frequencies align clearly in such a diagram due to their arithmetic relationships.  

In Figure \ref{fig:jumps}, the points on the right-hand side of the diagram and climbing upwards to the left with 
regular ``jumps'' are the linear combinations of $f_0$ and $f_1$, or $ i{\cdot} f_1 + j{\cdot}f_0$. (For this
diagram, they are written in the practical form of $ k{\cdot} f_1 + l{\cdot}f_0 + m (f_1 - f_0) $, where $i, j$ and 
$k, l, m$ are integer numbers.)  The ``jumps'' in the diagram occur when the linear combinations go up by one 
order, and are due to the simple fact that $f_1 \rm mod (f_1-f_0) = f_0 \rm mod (f_1-f_0) \neq 0$. Hence, for 
every additional $f_0$ or $f_1$ in a linear combination (besides multiples of $(f_1-f_0)$), the ``folded'' (modulo) 
value changes. The change corresponds to ${\delta}f = 3{\cdot} f_1 - 4{\cdot}f_0 = 0.03879\,d^{-1}$.  The 
points at almost the same y-value in the diagram show that, from this MOST data set, we were able to 
separate different linear combinations of $f_0$ and $f_1$ as close as 0.03879 d$^{-1}$ to one another.
The frequencies which fall in the left bottom corner of Figure \ref{fig:jumps} are associated with the additional 
frequencies at 1.96 and 3.92  d$^{-1}$. Note that those frequencies also follow a pattern of regular ``jumps'', 
since we have detected linear combinations of these frequencies with $f_0$ and $f_1$.  We stress that the 
regular patterns are merely a consequence of the arithmetic relations between the frequencies in the linear 
combinations.

The MOST photometry of AQ Leo opens the possibility of a detailed analysis of the properties of the 
combination frequencies resulting from nonlinear coupling, which can be used to test hydrodynamical 
models of RRd stars (see Section\,4.2). 

\begin{figure}
\includegraphics[width=8cm]{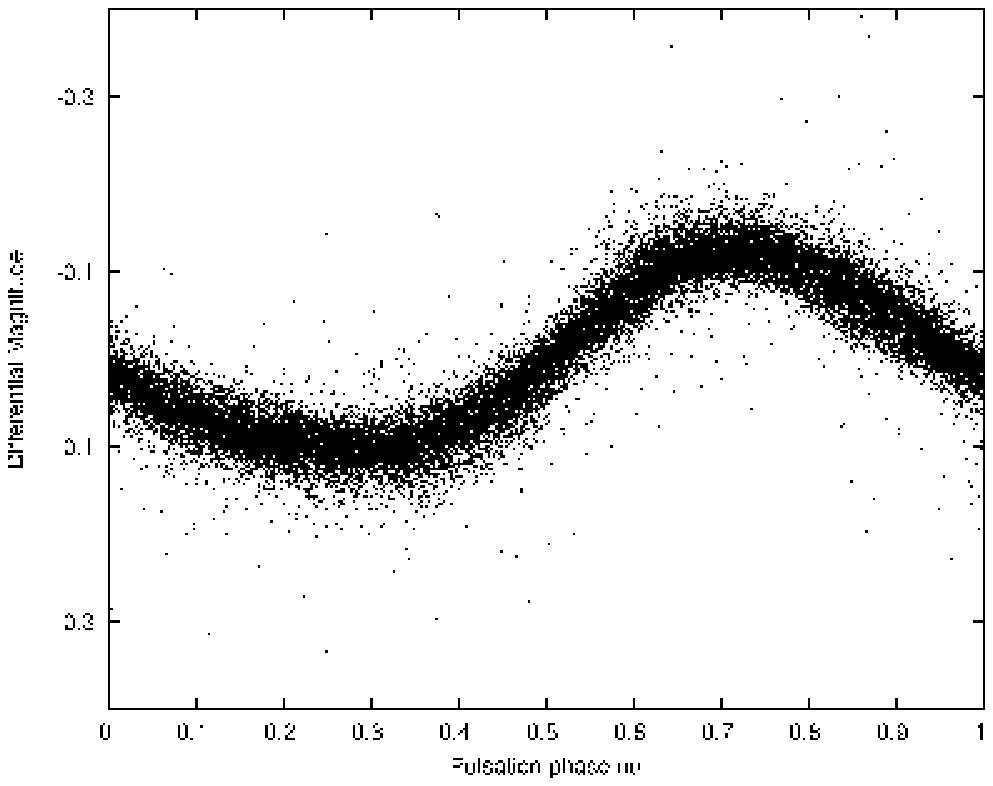}
\includegraphics[width=8cm]{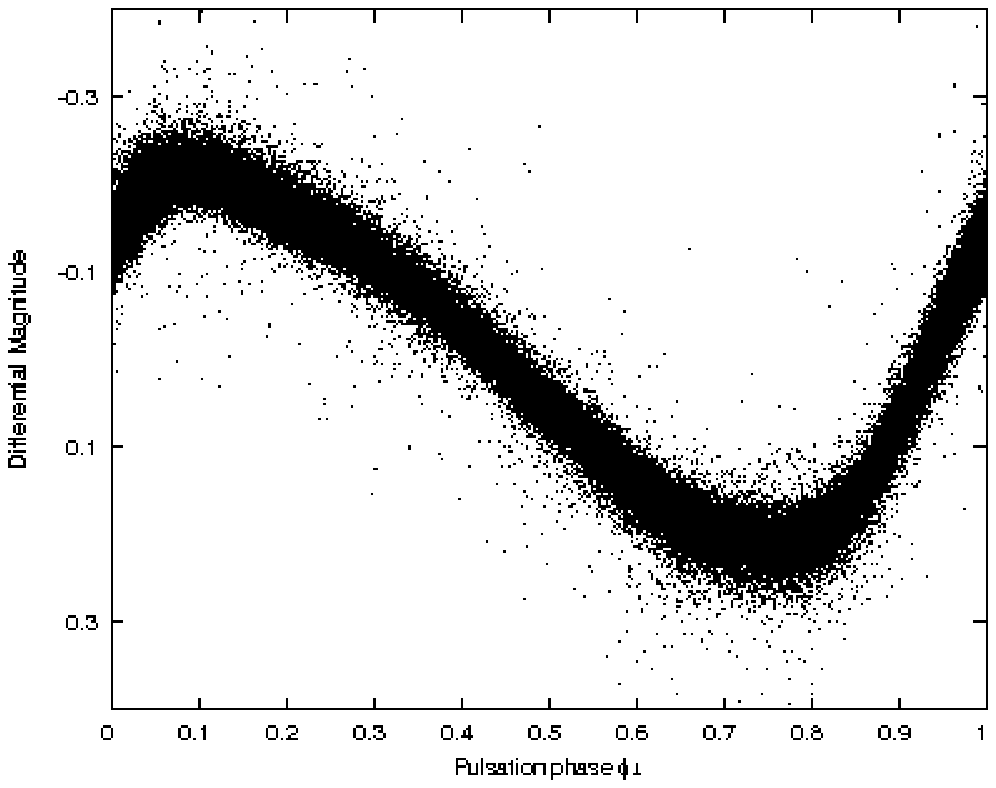}
\caption{The deconvolved residuals of the MOST light curve of AQ Leo (after removal of combination 
frequencies and the additional frequencies) folded with the fundamental radial mode period $P_0$ (upper 
panel) and with the first overtone radial mode period $P_1$ (lower panel).} 
\label{fig:phase}
\end{figure}

The phase diagrams of the two dominant modes are shown in Figure \ref{fig:phase}, in which the linear 
combinations and frequencies unrelated to the two dominant modes we find in the fit (Table \ref{tab:freqs}) 
have been subtracted.  In AQ Leo, the first overtone mode is the dominant one.  It is characterised by 
stronger nonlinear behaviour which can be seen by the more pronounced nonsinusoidal shape of its light 
curve in Figure \ref{fig:phase}.  

\subsection{Amplitude ratios and comparison to theory}

Early in the study of RRd stars, it was believed that they were in the process of switching from the fundamental 
radial mode to the first overtone or vice versa.  Cox et al. (1980) concluded that the double-mode behaviour 
of AQ Leo could be explained by mode switching during its blueward or redward horizontal branch evolution.  
This mode-switching phase is very brief compared to the star's duration as an RR Lyrae variable - lasting 
only a few thousand years (Cox et al. 1980, 1983). This is too short to account for the observed numbers of 
RRd stars in clusters like M15 (see Papar\'o et al. 1998), which suggests that the RRd stage must on 
average last longer than the theoretically predicted mode-switching time scales.  Feuchtinger (1998) was 
the first to find stable (i.e., not resulting from mode switching) double-mode behaviour in RR Lyrae models 
by including a time-dependent mixing length model for convection. Subsequently, convection was established 
by Kollath et al. (2002) as the key player in double-mode pulsation in Cepheid and RR Lyrae models . 

Even so, there is some observational evidence that the transition from a monoperiodic RR Lyrae star (RRc: 
redward, or RRab: blueward) to a double-mode RR Lyrae star can take place over a very short timescale. 
Clement \& Goranskij (1999) observed a rapid mode change from RRab to RRd in the star V79 in the cluster 
M3 in less than a year.  Buchler \& Kollath (2002) have shown that bifurcation points (characterized by 
changes in pulsation behaviour) exist where pulsation is no longer independent from stellar evolution. Taking 
this into account, the number of observed double-mode pulsators suggests that these oscillations are stable 
and that rapid mode switching is a relatively rare effect. Still, the true nature of double-mode pulsation in 
individual stars remains elusive from the observers' point of view.

If the kinetic energy switching times are only a few hundred years or less and the relative amplitude change
timescale less than a thousand years, the relative amplitudes of the two modes might be seen to change over 
timescales of decades. The amplitude ratio of the first overtone to fundamental mode in AQ Leo in 1973-1975 
measured by JW77 was $A_1/A_0 \simeq 2$. The amplitude ratio measured by MOST in 2005 is $A_1/A_0 
\simeq 1.8$. Unfortunately, it is risky to compare our observed amplitudes and phases to those tabulated by 
JW77, because the MOST photometry was obtained through a wide nonstandard passband while the JW77 
data were obtained through a Johnson $B$ filter.  

\begin{figure}
\includegraphics[width=8cm]{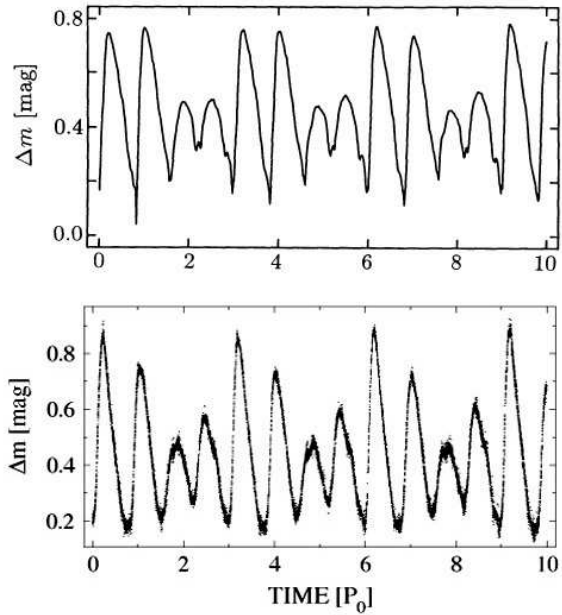}
\caption{Comparing light curves from a computer and from space. A segment of the theoretical light curve 
of a purely radiative RRd star, taken from Kov\'acs \& Buchler (1993) covering 10 cycles of the fundamental 
mode period in their model (upper panel - reproduced by permission of the AAS). A segment of the MOST 
light curve of AQ Leo spanning the same amount of time, shown at the same scale (lower panel).  It is now 
possible to obtain data which is sampled as thoroughly and with noise levels comparable to time-resolved 
model calculations.}
\label{fig:complc}
\end{figure}

\begin{figure}
\includegraphics[width=8cm]{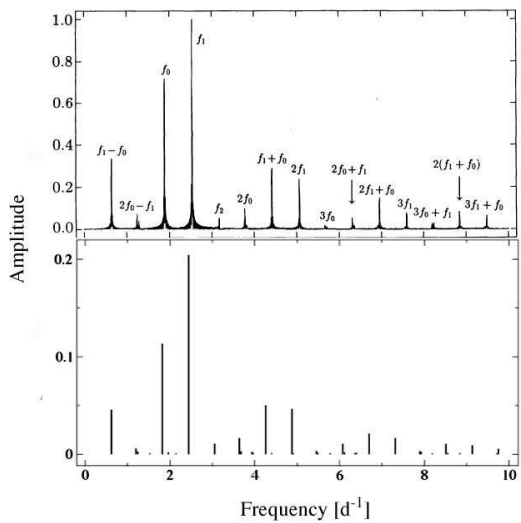}
\caption{Amplitude spectra of the model light curve from Fig.\,\ref{fig:complc}, taken from Kov\'acs \& Buchler 
(1993) (upper panel - reproduced by permission of the AAS)  and  the MOST photometry in the same 
frequency range (lower panel). The relative amplitudes compare quite well to our findings.} 
\label{fig:compspec}
\end{figure}

A qualitative comparison to a RRd model (see Kov\'acs \& Buchler 1993) is presented in Figures \ref{fig:complc} 
and \ref{fig:compspec}. The striking similarities to the model results in the MOST light curve and resulting 
amplitude spectrum shows that current spacebased observations are able to reach the quality of numerical 
simulations. Admittedly, the frequencies of the fundamental and the first overtone themselves don't match and 
we do not propose this model to be an accurate description of AQ Leo. The model was calculated under the 
unrealistic assumption of a purely radiative stellar envelope. By today's standards, more sophisticated
treatments of the hydrodynamic equations (including convection) in the calculations are more reasonable 
(Koll\'ath \& Buchler 2001). Still, the predicted relative amplitudes of the various linear combinations of $f_0$ 
and $f_1$ behave in close agreement with our observations up to very high order. As such, observations of the 
quality presented in this paper should be considered to be a tool of validation for the non-linear pulsational 
behaviour of modern RRd models.  The observed frequencies, their amplitudes and phases can even be a 
testing ground for the development of hydrodynamical models of RRd stars including nonradial pulsations.

\subsection{Period changes?}

Basic pulsation physics predicts that an increase in the fundamental mode period would be accompanied by 
an increase in the first overtone mode period, reflecting an increase in the mean stellar density. Papar\'o et 
al. (1998) studied period changes in RRd stars in M15, and found that the rate of period change for the 
fundamental radial mode is significantly larger than for the first overtone. For most of the stars in their 
sample, the ratio $P_0/P_1$ is increasing.  They found the period changes of the fundamental modes of the 
RRd stars were mostly negative, while for the first overtone, both positive and negative changes were seen.
The most striking result by Papar\'o et al. (1998) is the measurement of period changes of different sign for 
the fundamental and first overtone modes.  This suggests, as for the $\delta$ Scuti pulsators, that the period 
changes may not be simply linked to an RR Lyrae star's evolution in the HR diagram.  

The uncertainties in the frequencies in Table \ref{tab:freqs} are slightly too large, given that the MOST time 
series spans only about 34 days, to measure reliably the predicted evolutionary period changes in the 30 
years between the MOST and JW77 observing runs. But Jerzykiewicz, Schult \& Wenzel (1982) (hereafter 
JSW82) deduced that there was a sudden increase of the period of the first overtone of AQ Leo in the early 
1970s, by comparing re-evaluated JW77 photometry to archived data dating back to 1929.  It is possible 
that there have been other period changes in AQ Leo unrelated to the evolution of the mean density of the star. 

Therefore, we compared the periods from the MOST photometry to the values reported in JSW82.  (We cannot 
perform a meaningful $O-C$ analysis due to the complexities of the light curves.)  Our results yield evidence
for frequency changes in AQ Leo since the JW77 observations outside the 1$\sigma$\, uncertainties: 
fundamental mode, \[ \Delta f_0 = -0.00005 \pm 0.00002\,d^{-1}\] and first overtone, \[ \Delta f_1 = +0.00007 
\pm 0.00001\,d^{-1}.\]  While the change of the fundamental frequency is consistent with zero within the 
3$\sigma$ confidence limits, the non-zero positive change of the first overtone is highly significant.  
Interestingly, the value of $f_1$ we obtained from the MOST data matches that proposed in JSW82 to satisfy 
an $O-C$ diagram covering the data obtained from 1929 up to right before the abrupt period change in the 
1970s. Our results fit their findings to within the 1$\sigma$ uncertainties. This hints that the period change in 
the 1970's was a short-lived event.

Consequently, notwithstanding the higher uncertainties, we obtain a slightly different period ratio than that of 
JSW82, namely $P_1/P_0 = 0.74602\pm 0.00002$ compared to $0.746063 \pm 0.000003$. This does not 
significantly disagree with the theoretical result of Cox et al. (1980), who calculated that the fundamental mode 
period and the period ratio of 0.746 can be explained if AQ Leo has a homogeneous composition typical of 
Population II and a mass of $0.65\,M_{\odot}$. More recent calculations (Koll\'ath, private communication) 
indicate, however, that a mass $> 0.7\,M_{\odot}$ is likely to be more accurate . When compared to the 
results of Szab\'o, Koll\'ath \& Buchler (2004), our values match the sequence with $M = 0.71\,M_{\odot}$, 
$L = 60\,L_{\odot}$ and very low metallicity.

\subsection{New modes in AQ Leo?}

RR Lyrae stars have long been considered prototypes of radially pulsating stars. Two types of RR Lyrae stars 
show multiperiodic behaviour: Blazhko stars and double-mode stars, like AQ Leo. The former are characterised 
by the Blazhko Effect (see Kolenberg 2004 and Kov\'acs 2001), which is a periodic modulation of the amplitude 
and/or phase of the light curve on timescales of a few up to hundreds of days. In the Fourier spectra of such 
stars, additional frequencies occur close to the dominant frequency which is identified with a radial mode.  If 
these frequencies correspond to pulsation modes in the star, they must be nonradial modes.  Van Hoolst, 
Dziembowski \& Kawaler (1998) showed that there is a dense spectrum of nonradial modes which have lower
moments of inertia (and hence are more easily excited) in the vicinity of the radial mode frequencies in RR 
Lyrae stars.

The MOST observations of AQ Leo are the first in which additional frequencies besides the fundamental radial 
mode, the first overtone radial mode and their linear combinations are detected. The two additional frequencies 
are $f_i \simeq 1.96\,d^{-1}$ and $f_{ii} \simeq 3.92\,d^{-1}$. (Linear combinations of these frequencies 
with the known radial modes are also found in the data, see also  Figure \ref{fig:jumps}.)  The amplitude ratio 
of $f_i$ to $f_{ii}$ is 0.64 $\pm$ 0.08.

Judging from the period ratios alone, $f_{ii}$ ($\simeq 2f_i$) could be the third radial overtone (Zoltan Koll\'ath, 
private communication). If it is a strange mode (essentially an acoustic surface mode; see Buchler, Yecko \& 
Koll\'ath 1997), $f_i$ could then be a nonradial mode in a resonant (2:1) interaction with the third overtone.  
This scenario has been considered theoretically in studies of pulsational resonance (Buchler, Goupil \& 
Hansen 1997), supporting our argument. If we are indeed witnessing resonance of a non-radial and a radial 
mode, it is the higher frequency which must correspond to $\ell=0$ due to parity considerations. The 
frequencies match the resonance within the uncertainties (see Table\,1). In case of a mismatch, the resonance 
cannot be locked.  The fact that we see significant amplitudes at both frequencies may suggest that AQ Leo is 
in a fleeting state of mode transition.  Previous calculations by Buchler, Yecko \& Koll\'ath (1997) predicted that if strange 
modes are present, they should correspond to higher overtones than the third.  Our new frequencies are more 
consistent with the third overtone, so models matched more closely to AQ Leo are necessary to compare with 
the MOST observations.  

The frequency $f_i$ -- close to the radial fundamental mode -- is also reminiscent of the additional frequencies 
seen in Blazhko stars. While the nature of the additional frequencies in AQ Leo, as well as in Blazhko stars 
(see, e.g., Kolenberg et al. 2003), must still be thoroughly investigated, there is increasing evidence that some 
RR Lyrae stars pulsate "beyond radial modes" (e.g., Clement \& Rowe 2000; Kiss et al. 1999). Our observations 
represent the first detection of additional modes in an RRd star.

\section{Where do we go from here?}

Nearly continuous MOST photometry of AQ Leo spanning about 34.4 days and with high photometric precision 
have provided (1) accurate Fourier components of the known radial fundamental and first overtone modes in this 
star; (2) an unprecedented number of 32 combination frequencies of these modes; (3) evidence for non-evolutionary 
changes in the periods of the fundamental and (even more strongly) the first overtone mode; and (4) additional 
frequencies which can only be plausibly explained by other modes which have never been seen before in AQ Leo, 
or in any RRd star until now.  What are the next steps forward?  

Observationally, it will be very difficult -- perhaps impossible -- to obtain a light curve of longer time coverage and 
better precision than the MOST photometry presented here in the near future.  The star will not be accessible to 
the CoRoT and Kepler space missions, nor will groundbased photometric campaigns reach what MOST has 
achieved for AQ Leo for some time to come.  If AQ Leo is reobserved by MOST in the coming years, it could 
provide a reliable measurement of the change in amplitude ratio of the radial fundamental and overtone modes in 
the MOST bandpass, to test theories of mode switching and evolution in this star. Also, further evidence for the 
period changes may be established. 

New MOST photometry may lead to the detection of even more frequencies (depending on the behaviour of AQ
Leo) but it will not be possible to improve the frequency resolution of the current data set dramatically.  
Time-resolved high-resolution spectroscopy to look for evidence of nonradial pulsation in the spectral line 
profile shapes would be highly desirable for this star.  Unfortunately, AQ Leo is quite faint ($V \sim 12.6$), so 
spectroscopy of sufficient wavelength resolution and high signal-to-noise are beyond reach of instruments which 
could monitor the star at a sufficient cadence for many weeks.  

The immediate path to better understanding of AQ Leo is theoretical modeling of the frequency spectrum identified
in the existing MOST photometry. The potential of the combination frequencies to test hydrodynamical models can
now be fully exploited thanks to the very precise MOST data.

\section*{Acknowledgments}

We would like to thank Zoltan Koll\'{a}th and Geza Kov\'{a}cs for their valuable contributions to the discussion. 
MG, TK, WWW (project number P17580-N02), and KK (project number P17097-N02) have received financial support by the Austrian Fonds zur F\"orderung der wissenschaftlichen 
Forschung.  JMM, DBG, AFJM, SR, DS and 
GAHW acknowledge funding from the Natural Sciences \& Engineering Research Council (NSERC) Canada. RK 
is supported by the Canadian Space Agency (CSA).

\end{document}